\documentclass[preprint,showpacs,preprintnumbers,amsmath,amssymb]{revtex4}
\usepackage{graphicx}

\begin{document}
\def \ee {\varepsilon}
\thispagestyle{empty}
\title{
Calculation of the Casimir force between Ge test bodies
in different theoretical approaches
}

\author{G.~L.~Klimchitskaya
}

\affiliation{
North-West Technical University, Millionnaya Street 5,
St.Petersburg, 191065, Russia
}

\begin{abstract}
We calculate the Casimir force between a Ge plate and a Ge spherical
lens with neglected and included free charge carriers.
Computations with included carriers are performed using the
Drude, plasma, and diffusion models. It is shown that
the plasma and the Drude models lead to almost identical
computational results, while the results obtained using the
diffusion model are sandwiched between those obtained with
neglected charge carriers and with charge carriers
described by the plasma or Drude models.
\pacs{12.20.-m}
\end{abstract}

\maketitle

Using the Lifshitz formula and the proximity force approximation,
the Casimir force between the spherical lens of radius $R$ at
a height $a\ll R$ above a plate at temperature $T$ is given by
\begin{equation}
F(a,T)=\frac{k_B TR}{4a^2}
\sum_{l=0}^{\infty}{\vphantom{\sum}}^{\prime}
\int_{\zeta_l}^{\infty}y\,dy \left\{
\ln[1-r_{\rm TM}^2({\rm i}\zeta_l,y){\rm e}^{-y}]
+\ln[1-r_{\rm TE}^2({\rm i}\zeta_l,y){\rm e}^{-y}]\right\}.
\label{eq1}
\end{equation}
Here, $k_B$ is the Boltzmann constant,
$\zeta_l=4\pi a k_B Tl/(\hbar c)$ with $l=0,\,1,\,2,\,\ldots$
are the dimensionless Matsubara frequencies, and prime adds
a multiple 1/2 to the term with $l=0$. In the Lifshitz theory
the reflection coefficients are
\begin{equation}
r_{\rm TM}({\rm i}\zeta_l,y)=\frac{\varepsilon_ly-
\sqrt{y^2+(\varepsilon_l-1)\zeta_l^2}}{\varepsilon_ly+
\sqrt{y^2+(\varepsilon_l-1)\zeta_l^2}},
\qquad
r_{\rm TE}({\rm i}\zeta_l,y)=\frac{y-
\sqrt{y^2+(\varepsilon_l-1)\zeta_l^2}}{y+
\sqrt{y^2+(\varepsilon_l-1)\zeta_l^2}},
\label{eq2}
\end{equation}
\noindent
where $\varepsilon_l\equiv\varepsilon({\rm i}\omega_c\zeta_l)$,
$\omega_c=c/(2a)$.

The dielectric permittivity of intrinsic Ge with neglected
free charge carriers can be approximately represented in
the form \cite{1}
\begin{equation}
\varepsilon_l=\varepsilon_{\infty}+
\frac{\varepsilon_0-\varepsilon_{\infty}}{1+
\frac{\zeta_l^2\omega_c^2}{\omega_0^2}},
\label{eq3}
\end{equation}
\noindent
where $\varepsilon_{\infty}=1.1$, $\varepsilon_0=16.2$, and
$\omega_0=5.0\times 10^{15}\,$rad/s.
Substituting Eqs.~(\ref{eq2}) and (\ref{eq3}) into
Eq.~(\ref{eq1}),
and performing computations for a sphere of $R=15.10\,$cm radius
at $T=300\,$K, we obtain the values of the Casimir force in the
experimental configuration of Ref.~\cite{2} with neglected role
of charge carriers.
The force magnitudes at a few typical experimental separations
are listed in the column 2 of Table 1.

Thermodynamically and experimentally consistent application
of the Lifshitz theory to dielectric materials suggests that
one should
simply ignore the role of free charge carriers \cite{3}.
Intrinsic Ge is dielectric because its conductivity vanishes
with vanishing $T$.
However, there are different approaches in the literature
attempting to include free charge carriers into the model
of dielectric response when calculating the thermal Casimir
force between dielectrics.
The density of charge carriers in intrinsic Ge at $T=300\,$K
is estimated as $n_e=n_h=2.3\times 10^{13}\,\mbox{cm}^{-3}$
\cite{1,4}. Then the plasma frequencies of electrons and holes
can be formally calculated as
\begin{equation}
\omega_{p(e,h)}=\left(\frac{4\pi n_{e,h}e^2}{m_{e,h}}
\right)^{1/2}.
\label{eq5}
\end{equation}
\noindent
Here, $m_e=0.12m$, $m_h=0.21m$ are the effective masses, and
$m$ is the mass of an electron. This equation leads to
$\omega_{p(e)}=7.8\times 10^{11}\,$rad/s,
$\omega_{p(h)}=5.9\times 10^{11}\,$rad/s.

If one includes the role of free charge carriers by means of
the Drude model, the dielectric permittivity takes the form
\begin{equation}
\varepsilon_l^{\rm D}=\varepsilon_l+
\frac{\tilde{\omega}_{p(e)}^2}{\zeta_l[\zeta_l+
\tilde{\gamma}_{(e)}]}
+\frac{\tilde{\omega}_{p(h)}^2}{\zeta_l[\zeta_l+\tilde{\gamma}_{(h)}]},
\label{eq6}
\end{equation}
\noindent
where $\tilde{\omega}_{p(e,h)}={\omega}_{p(e,h)}/\omega_c$,
$\tilde{\gamma}_{(e,h)}={\gamma}_{(e,h)}/\omega_c$,
the relaxation parameters at $T=300\,$K are equal to
$\gamma_{(e)}\approx\gamma_{(h)}
\approx 2.6 \times10^{11}\,\mbox{s}^{-1}$ \cite{1},
and $\varepsilon_l$ is defined in Eq.~(\ref{eq3}).
It is easily seen that for all $l\geq 1$ the sum of the second
and third terms on the right-hand side of
Eq.~(\ref{eq6}) is less than $10^{-4}$\%
of  $\varepsilon_l$. Thus, charge carriers influence the
computational results only through the zero-frequency term
of the Lifshitz formula. The respective reflection
coefficients are
\begin{equation}
r_{\rm TM}^{D}(0,y)=1, \quad
r_{\rm TE}^{D}(0,y)=0.
\label{eq7}
\end{equation}
\noindent
The magnitudes of the Casimir force in the experimental configuration
of Ref.~\cite{2}, including the role of free charge carriers
described by means of the Drude model, are given in the third
column of Table 1.

Reference \cite{2} performs computations of the Casimir force not
only using the Drude model, but plasma model as well.
In this case the dielectric permittivity $\varepsilon_l^p$
is given by Eq.~(\ref{eq6}) with
$\tilde{\gamma}_{(e)}=\tilde{\gamma}_{(h)}=0$.
Note that for dielectric materials it is not justified to
describe free charge carriers by means of the plasma model
(the plasma model provides a consistent description of carriers
in the framework of the Lifshitz theory only for metals \cite{5}).
However, if one substitutes the dielectric permittivity of
the plasma model $\varepsilon_l^p$ in Eqs.~(\ref{eq1}) and
(\ref{eq2}) the magnitudes of the Casimir force shown in the
column 4 of Table 1 are obtained. As in the case of the
Drude model, charge carriers influence the results
that are obtained, only
through the term of the Lifshitz formula with $l=0$.
The respective reflection coefficients are
\begin{equation}
r_{\rm TM}^{p}(0,y)=1, \quad
r_{\rm TE}^{p}(0,y)=\frac{y-\sqrt{y^2+
\tilde{\omega}_{p(e)}^2
+\tilde{\omega}_{p(h)}^2}}{y+
\sqrt{y^2+\tilde{\omega}_{p(e)}^2
+\tilde{\omega}_{p(h)}^2}}.
\label{eq8}
\end{equation}
\noindent
{}From the comparison of columns 3 and 4 in Table 1 it follows
that for intrinsic Ge the Casimir forces computed using the
Drude and the plasma models are almost identical.
Thus the calculations of Ref.~\cite{2}, which resulted in significantly
larger magnitudes of the Casimir force computed using the plasma
model than those computed using the Drude model, are incorrect.

For comparison purposes, in column 5 of Table 1 we include also
the computational results obtained in the framework of the
diffusion model \cite{1,5}. As is seen in Table 1,
the magnitudes of the Casimir force computed using the
diffusion model are sandwiched between those computed with
neglected charge carriers, and the almost identical results
computed using the plasma or Drude models.


\begingroup
\squeezetable
\begin{table}[h]
\caption{\label{tab1} Magnitudes of the Casimir
force computed at $T=300\,$K
with neglected charge carriers (column 2)
and with charge carriers included in the framework of
the Drude (column 3),  plasma (column 4) and
 diffusion (column 5) models.}
\begin{ruledtabular}
\begin{tabular}{ccccc}
$a$&\multicolumn{4}{c}{$|F(a,T)|\,$(pN)}\\
\cline{2-5}
$(\mu\mbox{m})$&
(2) & (3) & (4) & (5) \\
\hline
0.6 & 679.22 & 748.03 & 748.11 & 706.63 \\
0.7 & 431.14 & 481.70 & 481.76 & 453.43 \\
0.8 & 291.28 & 329.99 & 330.05 & 309.79 \\
0.9 & 206.45 & 237.04 & 237.09 & 222.08 \\
1.0 & 152.00 & 176.78 & 176.82 & 165.39 \\
\end{tabular}
\end{ruledtabular}
\end{table}
\endgroup

\end{document}